\documentclass[a4paper]{jpconf}
\usepackage{graphicx}
\begin{document}
\title{Constraining in-medium heavy-quark energy-loss
	mechanisms via angular correlations between heavy and
	light mesons
	}

\author{M Rohrmoser, P-B Gossiaux, T Gousset, and J Aichelin}

\address{Subatech, 4 rue Alfred Kastler, 44307 Nantes, F}

\ead{martin.rohrmoser@subatech.in2p3.fr, gossiaux@subatech.in2p3.fr, gousset@subatech.in2p3.fr, aichelin@subatech.in2p3.fr .}

\begin{abstract}
	Two-particle correlations obtained from parton showers that pass the hot and dense medium of the quark gluon plasma (QGP)  can be used as an alternative observable, in addition to the combination of the nuclear modification factor $R_{AA}$ and the elliptic flow $v_2$, to study the mechanisms of in-medium heavy quark energy-loss. In particular, angular correlations represent a promising tool to distinguish between energy loss due to collisional and radiative interactions of jet and medium particles.
	To this end, parton cascades were created in Monte-Carlo simulations, where individual particles can undergo both parton splitting as well as an effective jet-medium interaction. 
	A first model simulates the effects of induced radiations on parton cascades. Its consequences on angular correlations of partons within jets were studied in detail, with particular focus on angular broadening. The results can be compared to a second model that effectively describes elastic scatterings of jet and medium particles.
\end{abstract}

\section{Introduction}
Energetic heavy quarks passing through the QGP, observed by the resulting mesons, are viewed as a suitable probe for the interactions inside the QGP, in particular for the mechanisms of energy loss, as they are less likely to thermalize within the medium and are mostly created at early stages of the medium evolution. However, models of both purely collisional energy loss and combinations of collisional and radiative energy loss are equally successful for reproducing the nuclear modification factor $R_{AA}$ and the elliptic flow $v_2$~\cite{Guiho}.

Angular correlations between two mesons represent alternative observables to make progress for discriminating between the two different interaction mechanisms. Previous studies, e.g. Ref.~\cite{nahrgang}, used azimuthal correlations between pairs of heavy mesons such as $D$-$\bar{D}$ pairs to distinguish between the energy-loss scenarios described. As an alternative, one can investigate the angular correlations between pairs of heavy and light mesons ($D$ and $\pi$) in the forward direction. This second approach is more sensitive to medium effects on the evolution of an individual jet, e.g. angular jet-broadening.

In order to perform such a study a Monte-Carlo code that simulates parton cascades, based on a similar approach as in Ref.~\cite{Renk2008}, was created. This algorithm allows for parton branchings and, furthermore, includes interactions of the cascade-particles with the medium in an effective model. 
In case of parton branchings, the outgoing partons and their four-momenta are selected from the distributions of splitting functions for collinear splitting, of leading order of $\alpha_s$, cf. Ref.~\cite{Altarelli1977298,Dokshitzer:1977sg}. 
The parton cascades are initiated by a single heavy quark with a certain energy $E_{ini}$, and a maximal virtuality $Q_\uparrow$ and follow an evolution down to a virtuality threshold $Q_\downarrow$. 

The angular two-particle correlations considered are distributions over the angles $\Delta \theta$ between the direction of the three-momenta of the heavy quark and that of every other cascade particle in its final state. Studies involving heavy quarks have the crucial advantage that the heavy particle can be tagged among the jet particles. From Ref.~\cite{Mele1991626} it can be deduced  that for sufficiently high virtualities $Q$, where $Q\gg m$, the radiation behavior of a heavy quark can be approximated by that of a massless quark. Therefore, the same splitting functions are used for, heavy as well as light quarks, while the explicit inclusion of a quark mass remains to be done in further studies.

\section{Model for inelastic jet-medium interaction}
The jet-medium interaction is described by a continuous increase of squared virtuality $Q^2(t)$ over time, according to the law $\dot{Q}^2(t)=\hat{q}(t)$, during the propagation of the partons in the medium, cf. Ref.~\cite{Renk2008}. This kind of approach serves as an effective model for medium-induced radiation, since partons with these additional amounts of virtuality will, on average, branch more often than their vacuum counterparts. In the Monte-Carlo implementation of the in-medium propagation, the continuous interactions with the cascade particles were discretized over small time-steps $\Delta t$, which leads to the following changes in parton virtuality over time:
$
Q(t+\Delta t)\mapsto\sqrt{Q^2(t)+\Delta t\, \hat{q}(t)}\,.
$
In order to avoid confusion of the medium effects present in cascades of this kind of effective model with consequences of three-momentum transfers from medium to jet particles, the virtuality increase was attributed solely to an increase in parton energy, i.e.:
$
E(t+\Delta t)\mapsto\sqrt{E^2(t)+\Delta t\, \hat{q}(t)}\,,\;
\vec{p}(t+\Delta t)\mapsto\vec{p}(t)\,.
$
The evolution of the transfer $\hat{q}(t)$ within the medium was fitted to a hydrodynamical description in Ref.~\cite{Renk2008}. In particular, it is assumed that the initial quarks of the cascades are created at the center of a medium that extends up to $L=10$~fm/c. In order to quantify the strength of the jet-medium interaction, the quantity $\Delta Q^2$ was used, defined as
$
	\Delta Q^2=\int_0^L\hat{q}(t)dt\,.
$

\begin{figure}[h!]
\includegraphics[scale=0.5,clip=true, trim=0 10pt 0pt 10pt]{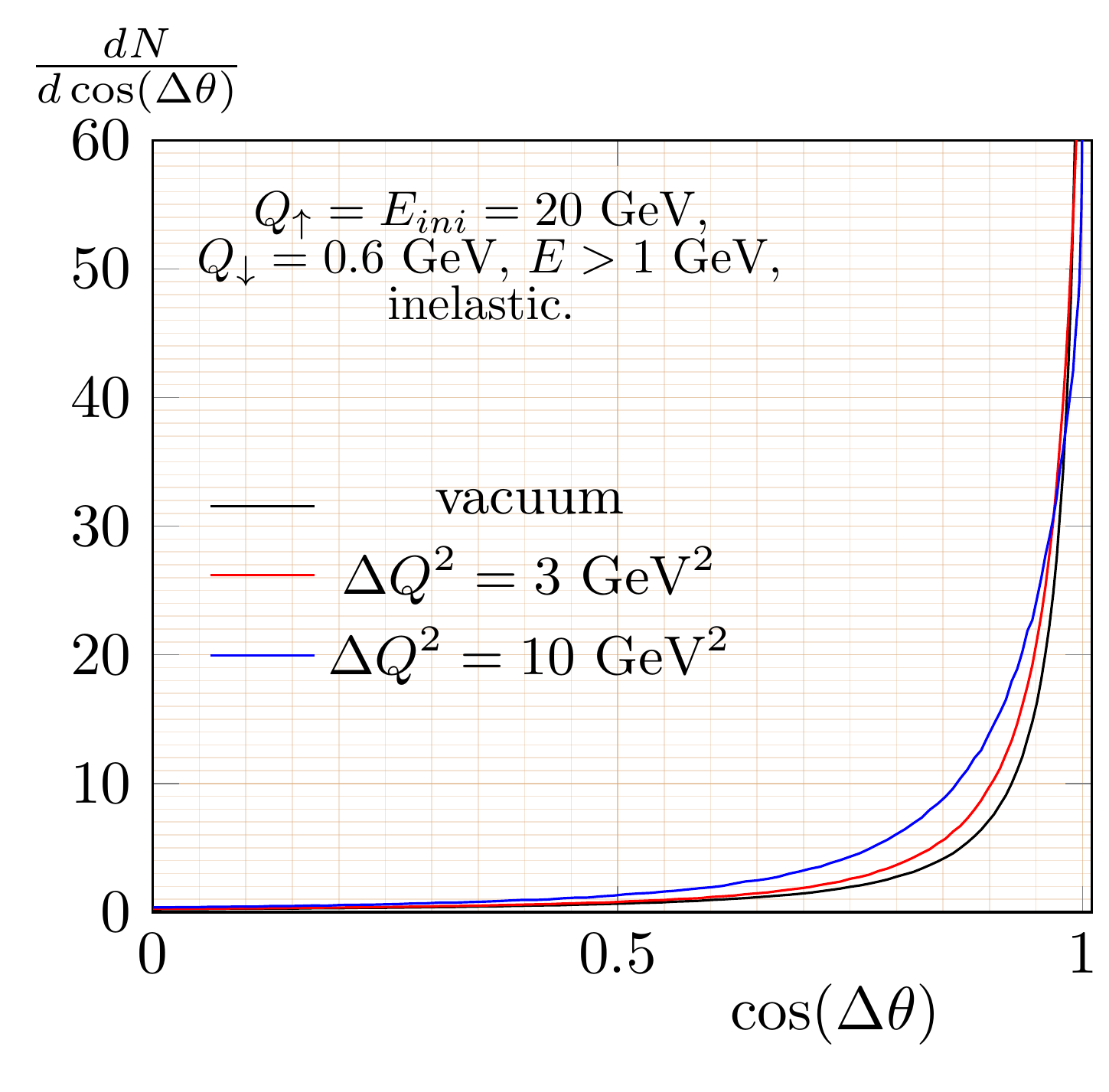}
\includegraphics[scale=0.5,clip=true, trim=0 10pt 0pt 10pt]{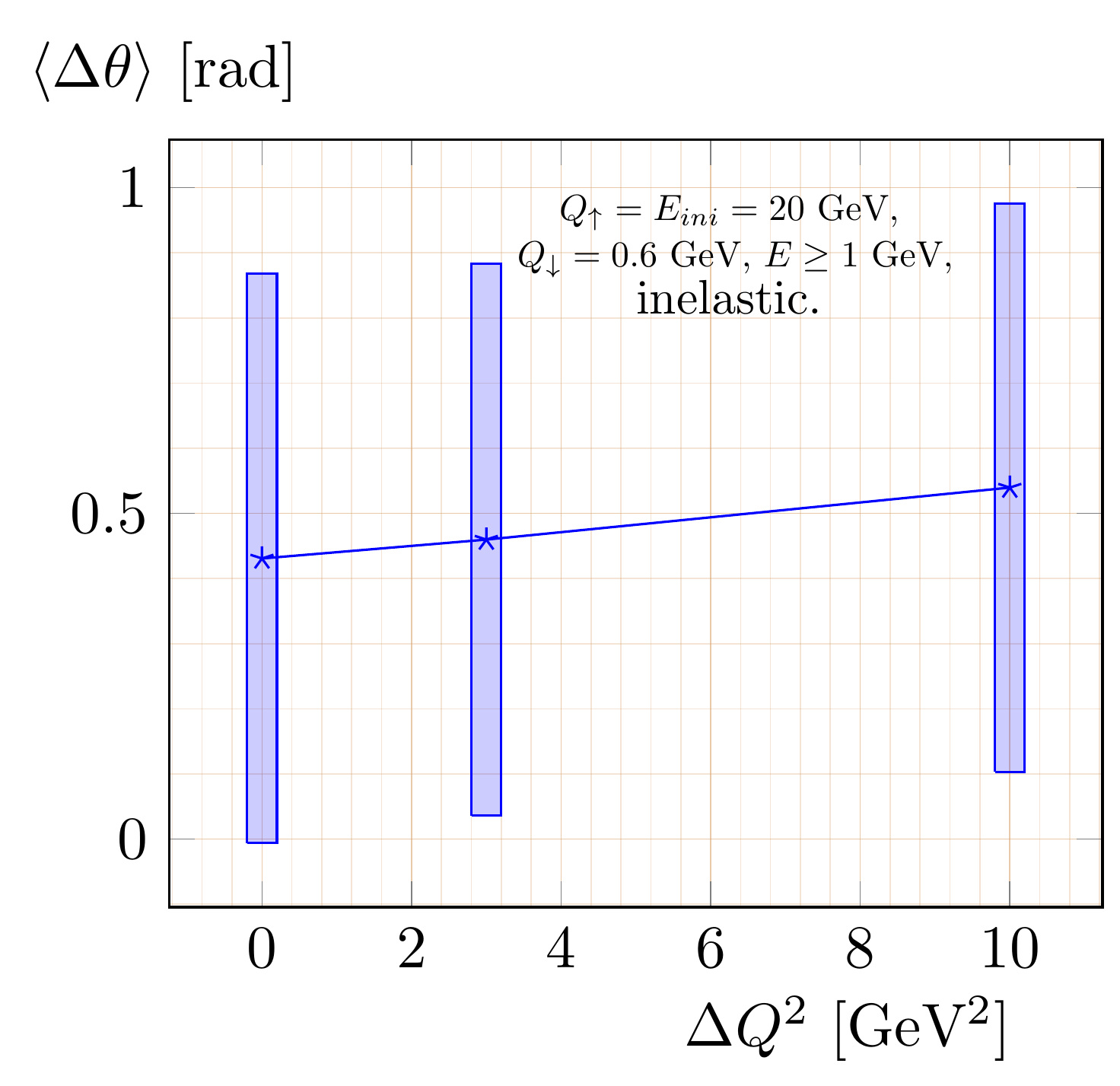}
\caption{left panel: angular correlations  
	in the form of $\frac{dN}{d\cos(\Delta \theta)}$, normalized to the average particle number per cascade. right panel: Average values of $\Delta\theta$ (stars) and corresponding standard deviation (boxes) for the three different angular correlations shown in the left panel.}\label{fig1}
\end{figure}

\section{Results for angular correlations}
The investigation of medium effects on the evolution of parton cascades by means of two-particle correlations focused initially on a verification of angular broadening, which is later on studied via more refined observables. The left panel of Fig.~\ref{fig1} shows angular correlations in the form of $\frac{dN}{d\cos(\Delta \theta)}$ for different values of $\Delta Q^2$. A broadening of the distribution with increasing $\Delta Q^2$ can be observed. In order to quantify this medium effect, the mean values of $\Delta\theta$ are depicted in the right panel of Fig.~\ref{fig1} as functions of $\Delta Q^2$. 

\begin{figure}[h!]
	\centering
	\includegraphics[scale=0.49,clip=true, trim=0 20pt 0pt 15pt]{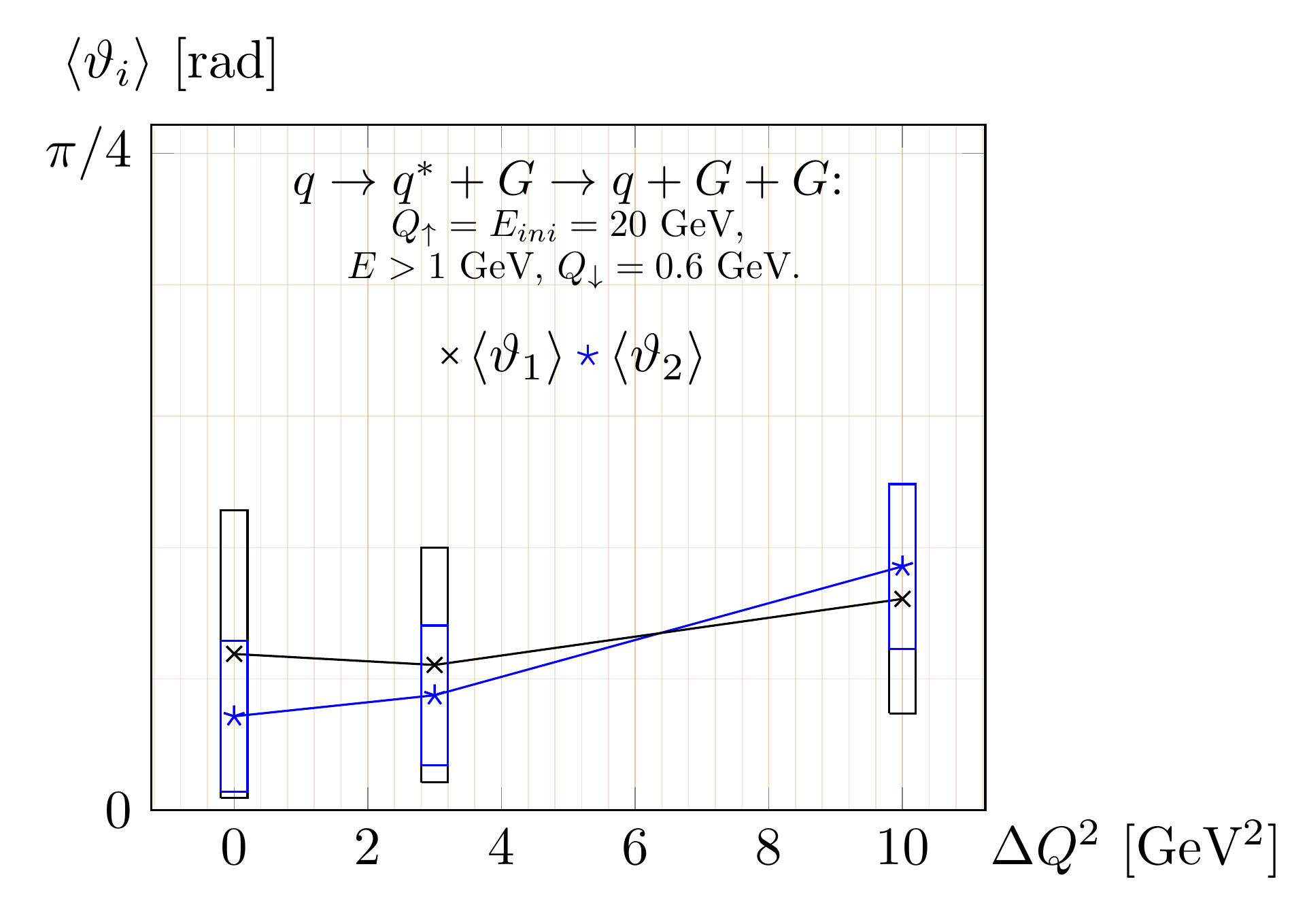}
	\caption{Average values of the first and second branching angles, $\vartheta_1$ (crosses) and $\vartheta_2$ (stars), respectively, as a function of $\Delta Q^2$ for cascades with two consecutive gluon radiations from a single quark line together with the standard deviations of the distributions over $\vartheta_1$ and $\vartheta_2$ (boxes). The cascades were extracted from the set used for the results in Fig.~\ref{fig1}.}\label{fig2}
\end{figure}

To understand how angular correlations are built up during jet evolution, the sets of parton cascades obtained by Monte-Carlo simulations were sorted with regard to their topologies. A simple interesting case consists of a quark that splits two times successively into a quark and a gluon. The average values of the angles between the outgoing quark and gluon in the first and second splitting, $\vartheta_1$ and $\vartheta_2$, respectively, are shown in Fig~\ref{fig2}. In contrast to the first branching angle the second branching angle increases strongly with $\Delta Q^2$. This behavior can be explained with the much smaller virtualities of the quark after the first splitting than in the initial state. Thus, the time between the first and second splitting is longer than the time before the first splitting, which results in higher amounts of accumulated virtuality from jet-medium interactions.

\begin{figure}[h!]
	\includegraphics[scale=0.48,clip=true, trim=0 25pt 0pt 15pt]{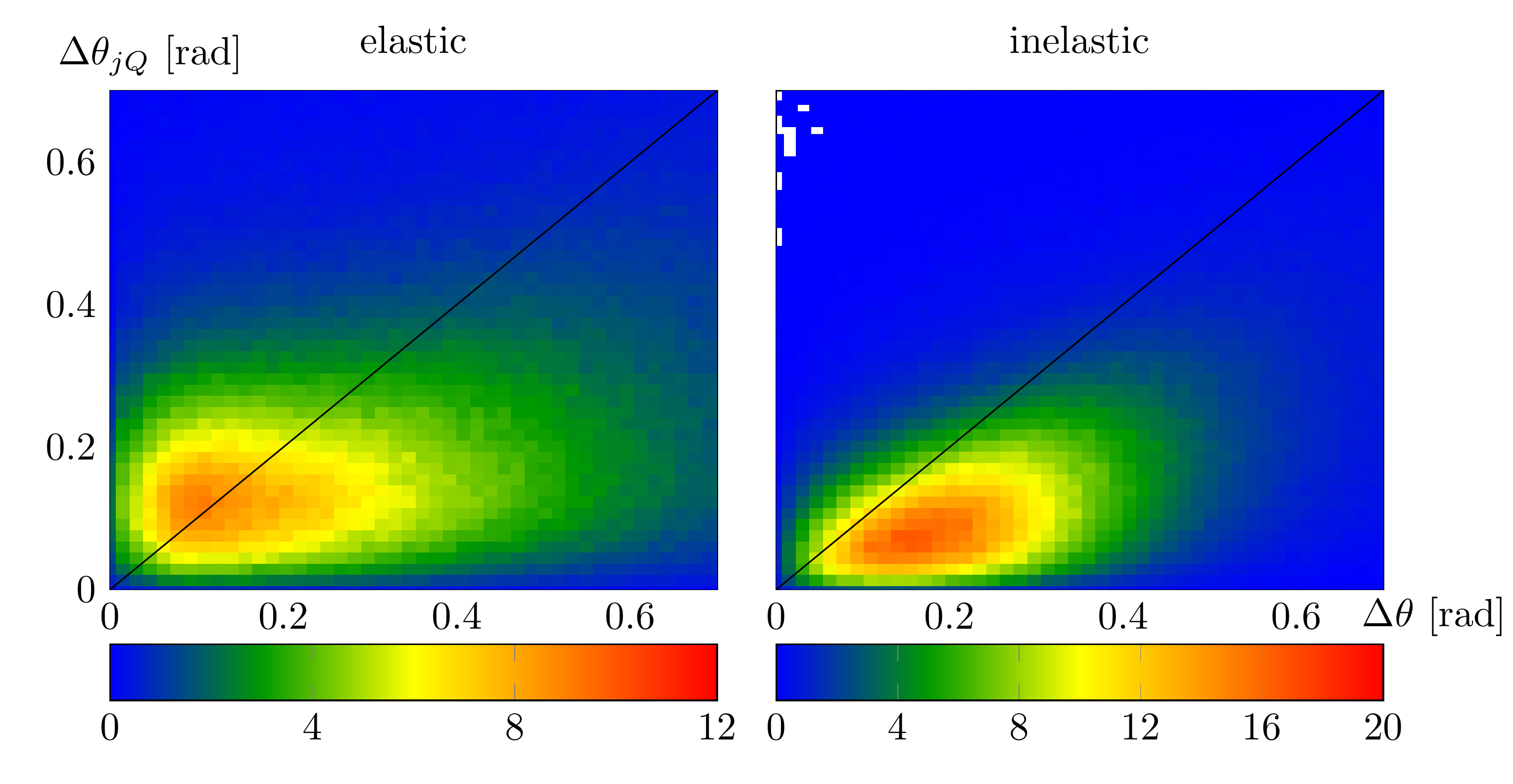}
	\caption{Distributions $\frac{d^2N}{d\Delta \theta_{jQ}d\Delta \theta}$ for parton cascades with $Q_\uparrow=E_{ini}=20$~GeV and $Q_\downarrow=0.6$~GeV that are subjected to elastic (left panel) and inelastic (right panel) jet-medium interactions, both for $\int_0^L\hat{q}dt=10$~GeV$^2$. Only particles with a three-momentum that fulfills $\|\vec{p}\|\geq2$~GeV were considered. The distributions are normalized to the average number of the corresponding parton pairs per cascade.}
		\label{fig3}
\end{figure}

Furthermore, in Fig.~\ref{fig2}, one observes angular ordering in the form $\vartheta_1>\vartheta_2$ in the vacuum. In the medium, this behavior is weakened, and for $\Delta Q^2=10$~GeV$^2$ even inversed. 
In order to find realistic observables that allow to look for such a kind of effect, and, furthermore, to discriminate between elastic and inelastic jet-medium interactions, the distribution $\frac{d^2N}{d\Delta \theta_{jQ}d\Delta \theta}$ was considered. There, $\Delta \theta_{jQ}$ is the angle between the directions of the three-momenta of the entire jet and of the tagged heavy quark. For the elastic jet-medium interactions, an effective model analogous to the one for inelastic scattering was used: In this model, however, the virtualities $Q$ are constant during in-medium propagation, while the partons experience a longitudinal drag force and transverse momentum transfers following the description in Ref.~\cite{Berrehrah:2014kba}. In the elastic model, the transport coefficient is $\hat{q}=\frac{d\langle p_T^2\rangle}{dt}$. Fig.~\ref{fig3} compares the distributions $\frac{d^2N}{d\Delta \theta_{jQ}d\Delta \theta}$ for both models with the same value $\int_0^L\hat{q}dt=10$~GeV$^2$. One can observe a strong broadening in $\Delta \theta$ for the elastic model, while this effect is less apparent for the inelastic model. For the latter, the peak of the distribution can be found at larger values of $\Delta \theta$ than for the elastic model, where the peak of the distribution occurs at higher $\Delta \theta_{jQ}$ values.  
\section{Conclusions and outlook}
It was verified for parton cascades produced in an effective model for medium-induced radiation that angular two-particle correlations are sensitive observables to an angular broadening in the medium. A further investigation of cascades of particular topologies showed that later stages of jet evolution are more sensitive to this medium effect than earlier ones, which ultimately results in an inversion of angular ordering. 

Currently, angular two-particle correlations are investigated in another effective model that is based on elastic collisions in an approach presented in Ref.~\cite{Berrehrah:2014kba}. 
It was shown that more sophisticated correlations could be useful to discriminate between the models for HQ-jet modifications by the medium, although they are more difficult to extract from the background.

\ack
M R thanks the organizers of the Strangeness in Quark Matter 2016 conference in Berkeley, CA, USA, and the project TOGETHER of the Region Pays de la Loire, France for their financial support that allowed for the participation in the conference.

\section*{References}

\end{document}